\documentstyle[prd,aps,preprint,tighten,floats]{revtex}
\begin{document}
\draft
\preprint{UPR-675-T, NSF-ITP-95-74}
\date{\it August 1995}
\title{BPS Saturated and Non-Extreme States in Abelian Kaluza-Klein
Theory and Effective $N=4$ Supersymmetric String Vacua
\thanks{Contribution to the Proceedings of {\it Strings'95} conference.}}
\author{Mirjam Cveti\v c
\thanks{E-mail address: cvetic@cvetic.hep.upenn.edu}$^{1,2}$
and Donam Youm\thanks{E-mail address: youm@cvetic.hep.upenn.edu}$^1$}
\address {$^1$ Department of Physics and Astronomy \\
          University of Pennsylvania, Philadelphia PA 19104-6396\\
          and\\  $^2$ Institute for Theoretical Physics,\\
          University of California, Santa Barbara, CA 93106-4030}
\maketitle
\begin{abstract}
{We summarize results for all four-dimensional
Bogomol'nyi-Sommerfield-Prasat (BPS) saturated and non-extreme solutions
of the ($4+n$)-dimensional Abelian Kaluza-Klein theory.
Within effective $N=4$ supersymmetric string vacua, parameterized in
terms of fields of the heterotic string on a six-torus, we then present
a class of BPS saturated states and the corresponding non-extreme
solutions, specified by $O(6,22,Z)$ and $SL(2,Z)$ orbits of general
dyonic charge configurations with zero axion.
The BPS saturated states with non-negative $O(6,22,Z)$ norms for
electric and magnetic charge vectors, along with the corresponding set of
non-extreme solutions, are regular with non-zero masses.  BPS saturated
states with the negative charge norms are singular, unaccompanied by
non-extreme solutions and become massless at particular points of the
moduli space.  The role that such massless states may play in the
enhancement of non-Abelian gauge symmetry as well as local
supersymmetry is addressed.}
\end{abstract}

\section{Introduction}

Solitons, {\it i.e.}, time-independent solutions of classical equations,
which saturate the Bogomol'nyi bound for their energy, shed light
on non-perturbative phenomena in non-linear field theories.
Even more intriguing is a recent recognition \cite{HTI,WITTENII} that
such configurations, also referred to as Bogomol'nyi-Sommerfield-Prasat
(BPS) saturated states, play a crucial role in addressing the full,
non-perturbative dynamics of string theory.  In particular, at points
of moduli space when such configurations become light they can affect
the low energy dynamics of the string theory in an important way
\cite{Hull,STROM,HTII}.  Thus, the study of BPS saturated states
for different string vacua may in turn shed light on the non-perturbative
string dynamics, as well as contribute to gathering
evidence for string-string duality between certain strongly coupled
and the corresponding weakly coupled string vacua.
In addition along with the BPS saturated states one would also like to
obtain information on the spectrum of the corresponding non-extreme
solutions, {\it i.e.}, configurations in the same topological class,
which are compatible with the corresponding Bogomol'nyi bound.
Although the latter set of states is in general modified by quantum
corrections, they may be relevant in the full string dynamics as well.

In this contribution we report on results for four-dimensional
BPS saturated states and the corresponding non-extreme solutions
which arise in effective supergravity theories compactified down
to four dimensions on tori, {\it i.e.}, on manifolds with Abelian
isometry.  In particular, we would like to shed light on the roles that
such states play in the effective $N=4$ superstring vacua.
Such string vacua are conjectured to be self-dual, {\it i.e.},
the string vacua of the heterotic string compactified on a six-torus ($T^6$)
transform into each other under the $SL(2,Z)$ transformations.
In addition, the heterotic string compactified on $T^6$ is
conjectured to be dual to the type IIA string compactified on a
$T^2 \times K3$ surface, which has its origin in the string-string
duality conjecture \cite{DUFFSS,WITTENII,HTI,STRDUAL} of the heterotic
and the type IIA string theories in six dimensions, as well as
to the eleven-dimensional supergravity compactified on $T^3\times K3$
surface, which has its origin in the duality conjecture
\cite{HTI,TOWNSEND,WITTENII} of the type IIA string theory and
the eleven-dimensional supergravity in ten dimensions.

At the conference, the results for all the four-dimensional static,
spherically symmetric BPS saturated states \cite{SUPER} as well as all
the non-extreme solutions \cite{CYALL} in the Kaluza-Klein sector
of the ($4+n$)-dimensional (minimally extended) supergravities
compactified on  $n$-tori have been presented.  The explicit form
of these solutions allows for a synthetic classification of all
of them.  For $n=7$, those are states in the Kaluza-Klein sector
of toroidally compactified eleven-dimensional supergravity
\cite{CYELEVEN} with $N=8$ supersymmetry in four dimensions,
which plays an important role as a dual theory \cite{HTI,WITTENII} of the
strongly coupled type IIA superstring theory on a six-torus.  For $n=6$,
those are states in the Kaluza-Klein sector of the toroidally compactified
heterotic sting theory \cite{CYKKHET}, which is dual to the type IIA
superstring on a $T^2\times K3$.

In this contribution, we also report on our recent work \cite{CYHET,CYHETS}
on a class of BPS saturated states of four-dimensional effective $N=4$
supersymmetric string vacua, which we parameterize in terms
of fields of the effective heterotic string theory compactified on a
six-torus.  We present BPS saturated states corresponding to
$O(6,22,Z)$ and $SL(2,Z)$ orbits of dyonic configurations with
zero axion; the $O(6,22,Z)$ orbits correspond to states with
the left-moving and the right-moving electric and magnetic charges
orthogonal, {\it i.e.}, light-like in the $O(6,22,Z)$ sense.
The states with the $O(6,22,Z)$ norms for the electric and magnetic
charges non-negative \cite{CYHET} correspond to regular solutions
with non-zero masses everywhere in the moduli space, while states with the
charge norms negative \cite{CYHETS} are singular solutions that become
massless (along with an infinite tower of states, related by $SL(2,Z)$
transformations) for  particular charge configurations and at
particular points of the moduli space.  Potential physical implications
of such massless states are also discussed.

We also address non-extreme solutions with the same charge content as
the  BPS saturated states discussed above.  The regular BPS saturated
states are accompanied by a set of regular non-extreme solutions
with masses compatible with the corresponding Bogomol'nyi bound.
On the other hand, singular BPS saturated states, which can become
massless at certain points of moduli space, have {\it no}
non-extreme solutions that are compatible with the
corresponding Bogomol'nyi bound.

In chapter 2, we give the results for static, spherically symmetric
solutions in the Abelian Kaluza-Klein theory with the most general
charge configurations, and discuss their thermal properties and
singularity structures.  In chapter 3, the general BPS saturated states
as well as the corresponding non-extreme ones for the effective theory
of $N=4$ superstring vacua are discussed.  Conclusions and open
problems are relegated to chapter 4.

\section{Spherically Symmetric Black Holes in Abelian Kaluza-Klein Theory}

The effective Abelian Kaluza-Klein theory \cite{KAL} in four dimensions
is obtained from ($4+n$)-dimensional pure gravity by compactifying
$n$ spatial coordinates on an $n$-torus by using the following Ansatz for
the ($n+4$)-dimensional metric:
\begin{equation}
g^{(4+n)}_{\Lambda \Pi}
=\left [ \matrix{{\rm e}^{-{1 \over \alpha}\varphi}g_{\lambda \pi} +
{\rm e}^{{2\varphi} \over {n\alpha}}\rho_{ij}A^i_{\lambda} A^j_{\pi} &
{\rm e}^{{2\varphi} \over {n\alpha}}\rho_{ij}A^i_{\lambda}
\cr {\rm e}^{{2 \varphi} \over {n\alpha}}\rho_{ij}A^j_{\pi} &
{\rm e}^{{2\varphi} \over{n\alpha}}\rho_{ij}} \right ],
\label{kkansatz}
\end{equation}
where $A^i_{\mu}$ is the Kaluza-Klein $n$ $U(1)$ gauge fields
with the field strengths $F^i_{\mu\nu}$, and the internal metric
$g_{ij}={\rm e}^{{2\varphi}\over {n\alpha}}\rho_{ij}$
(with $\rho_{ij}$ its unimodular part) and the four-dimensional
metric $g_{\mu\nu}$ depend on the four-dimensional
space-time coordinates, only.  We use the mostly positive signature
convention $(+++-+\cdots +)$ for $g^{(4+n)}_{\Lambda\Pi}$ with the
time coordinate in the fourth place, and $\alpha=\sqrt{{n+2}\over n}$.
The four-dimensional effective action (see for example
Ref. \cite{STAT}) has the global $SO(n)$ target space symmetry:
\begin{equation}
\rho_{ij} \rightarrow U_{ik} \rho_{k\ell} (U^{T})_{\ell j}, \ \ \ \ \ \ \
A^i_{\mu} \rightarrow U_{ij} A^j_{\mu},
\label{son}
\end{equation}
where $U$ is an $SO(n)$ rotation matrix, as well as the rescaling
symmetry \cite{STAT}.  In addition, for static or stationary
four-dimensional configurations the time-translation can be
considered, along with $n$ parameters of the internal isometry,
as a part of the $(n+1)$-parameter Abelian isometry group of a
($4+n$)-dimensional space-time manifold $\bf{M}$.  In this case the
projection of the ($4+n$)-dimensional manifold $\bf M$ onto the set
$\bf S$ of the orbits of the isometry group in $\bf M$ allows one to
express the ($4+n$)-dimensional Einstein gravity action in the
following three-dimensional one \cite{DM}:
\begin{equation}
{\cal L}= -{1 \over 2}\sqrt{-h}[{\cal R}^{(h)} -
{1\over 4}{\rm Tr}(\chi^{-1}\partial_a \chi \chi^{-1} \partial^{a}\chi)],
\label{threelag}
\end{equation}
where $h_{ab} \equiv \tau g^{\perp}_{ab}$ ($a,b=1,2,3$) is the
rescaled metric on $\bf S$ and
\begin{equation}
\chi = \left [ \matrix{\tau^{-1} & -\tau^{-1}\omega^{T} \cr
-\tau^{-1}\omega & \breve{\lambda} + \tau^{-1} \omega \omega^{T}}\right ]
\label{scalar}
\end{equation}
is the $(n+2) \times (n+2)$ symmetric, unimodular matrix of scalar
fields on $\bf S$.  Here, $\breve{\lambda}_{ij} \equiv g^{(4+n)}_{\Lambda\Pi}
\xi^{\Lambda}_i \xi^{\Pi}_j$, $\tau \equiv {\rm det}\breve{\lambda}_{ij}$
and $g^{\perp}_{ab} \equiv g^{(4+n)}_{ab} - \breve{\lambda}^{ij}
\xi_{ia}\xi_{jb}$.  The ``potential'' $\omega^{T} \equiv
(\omega_1,...,\omega_{n+1})$ defined as $\partial_a \omega_i =
\omega_{ia} \equiv \hat{\epsilon}_{abc}\xi^{b;c}_i$ ($\hat{\epsilon}_{abc}
\equiv \epsilon_{abc4...(4+n)}$) replaces the degrees of freedom of
$\xi_{ia}=g^{(4+n)}_{i+3,a}$.  The effective three-dimensional
Lagrangian density (\ref{threelag}) is invariant under the global
$SL(2+n,R)$ target space transformations\cite{DM}:
\begin{equation}
\chi \rightarrow {\cal U}\chi {\cal U}^{T},\ \ \ \ \ \
h_{ab} \rightarrow h_{ab},
\label{sltran}
\end{equation}
where ${\cal U} \in SL(2+n,R)$.

For the purpose of obtaining  static, spherically symmetric
configurations, one chooses the following Ansatz for the
four-dimensional metric:
\begin{equation}
g_{\mu\nu}dx^{\mu}dx^{\nu} =-\lambda(r)dt^2 + \lambda^{-1}(r)dr^2
+ R(r)(d\theta^2 + {\rm sin}^2 \theta d\phi^2).
\label{sphfourmet}
\end{equation}
The corresponding three-dimensional metric is then given by $h_{ab} =
{\rm diag}(1,\lambda R, \lambda R \,{\rm sin}^2 \theta)$ and
$n$ $U(1)$ gauge fields solve the equations of motion with
spherically symmetric Ans\" atze for the other fields, while the scalar
fields associated with the internal metric depend on the radial
coordinate $r$, only.

\subsection{Supersymmetric Configurations}

Among a class of solutions with a given charge configuration the
solution that saturates the corresponding Bogomol'nyi bound, referred
to as BPS saturated states, corresponds to the minimum energy (or vacuum)
configuration in its class.  These BPS saturated states satisfy the
Killing spinor equations which are obtained by setting the
supersymmetric variations of fermionic fields equal to zero, and
therefore are bosonic configurations which preserve some of the
supersymmetries.

With the Kaluza-Klein Ansatz for the ($4+n$)-dimensional metric
(Eq. (\ref{kkansatz})) one turns off all the other bosonic fields
of the corresponding supergravity theory.
Then the only non-trivial Killing spinor equations \cite{SUPER} turn
out to be those arising from the vanishing of the supersymmetry
transformation of (dimensionally reduced) ($4+n$)-dimensional gravitini.

With the spherical Ans\" atze for the four-dimensional fields,
the Killing spinor equations (corresponding to the $t$, $\theta$ and
$\phi$ components of the four-dimensional gravitini)
restrict a general supersymmetric configuration in this
class to have $n$ electric and $n$ magnetic charges subject to the
following orthogonality constraint \cite{SUPER}:
\begin{equation}
\vec{Q} \cdot \vec{P} = 0.
\label{kkchcon}
\end{equation}

All the supersymmetric configurations in this class can therefore
be obtained by imposing $SO(n)/SO(n-2)$ rotations (with $2n-3$
parameters) on the supersymmetric solution with
$U(1)_M \times U(1)_E$ charge configuration \cite{STAT}.  The latter one,
which we refer to as the generating configuration, is parameterized by
one magnetic charge $P$ and one electric charge $Q$, arising from
{\it different} $U(1)$ factors. It turns out that for this charge
configuration, among the scalar fields only
diagonal components of internal metric $g_{mn}$ are turned on
\footnote{In general with a diagonal internal metric Ansatz
the static, spherically symmetric configurations can have at most
one electric and one magnetic charge \cite{STAT},
which can also arise from the same $U(1)$ gauge fields.}.
The generating solutions with only electric [or only magnetic] charge
turned on preserve $1\over 2$ of the original supersymmetry while the
dyonic ones preserve $1\over 4$ of the original supersymmetry \cite{SUPER}.

Four-dimensional space-time for such
solutions is specified by two parameters $|\vec{P}|$ and $|\vec{Q}|$,
with  the following ADM mass:
\begin{equation}
M_{\rm BPS}=|{\vec P}|+|{\vec Q}|.
\label{bogobound}
\end{equation}
The explicit form of these solutions corresponds to a special case of
the general class of solutions with the ADM masses compatible with
the Bogomol'nyi bound (\ref{bogobound}) and are discussed in the
following subsection.

\subsection{General Class of Configurations}

General, four-dimensional, static, spherically symmetric solutions
are parameterized by the ADM mass $M\ge M_{\rm BPS}$, $n$ electric
 $\vec{Q} = (Q_1,...,Q_n)$ and $n$ magnetic $\vec{P} = (P_1,...,P_n)$
charges.  The explicit solution with such configurations was obtained
\cite{CYALL} by performing symmetry transformations (of the
three-dimensional action (\ref{threelag})) on the Schwarzschield solution.

By using the $SO(n)$ and the rescaling symmetry of the corresponding
four-dimensional action, one can bring the asymptotic value of
matrix $\chi$ (Eq.(\ref{scalar})) into the form
$\chi_{\infty} = {\rm diag}(-1,-1,1,...,1)$.  Then, the subset of
$SL(2+n,R)$ symmetry transformations of three-dimensional action
Eq. (\ref{threelag}) that preserves this asymptotic form of $\chi$ is
$SO(2,n)$.

By performing a set of two $SO(1,1)$ boosts (on the 1$st$ and the
($n+1$)-$th$, and the 2$nd$ and the ($n+2$)-$th$ indices of $\chi$)
with the boost parameters $\delta_{P,Q}$ on the Schwarzschield
solution, {\it i.e.}, $\chi ={\rm diag}(-(1-{m\over r})^{-1},
-(1-{m\over r}),1,...,1)$, one obtains the non-extreme
$U(1)_M \times U(1)_E$ black hole solutions parameterized
in terms of the magnetic  charge $P \equiv m{\rm sinh}
\delta_P {\rm cosh} \delta_P$, electric  charge $Q = m{\rm sinh}
\delta_Q {\rm cosh} \delta_Q$, and the ADM mass
$M\equiv m({\rm cosh}^2 \delta_Q + {\rm cosh}^2 \delta_P )$.
The ADM mass is traded for the non-extremality
parameter $\beta \equiv {m\over 2}>0$.
Additional two $SO(1,1)$ boosts (on the 1$st$ and the ($n+2$)-$th$, and
the 2$nd$ and the ($n+1$)-$th$ indices of $\chi$) with boost parameters
$\delta_1$ and $\delta_2$ generate a solution with the following
explicit form \cite{CYALL}:
\begin{eqnarray}
\lambda&=&{{r(r+2\beta)} \over {(XY)^{1/2}}},\
R=(XY)^{1/2},\   e^{{2\varphi} \over \alpha}=
{X \over Y},\  \rho_{ij}=\delta_{ij}{\rm e}^{-{{2\varphi}\over{n\alpha}}}
\ \ (i,j \neq n-1,n),
\nonumber\\
\rho_{n-1,n-1}&=&{{We^{{{2(n-2)}\over {n\alpha}}\varphi}} \over
{(XY)^{1/2}}}, \ \
\rho_{n-1,n}={{Ze^{{{2(n-2)}\over{n\alpha}}\varphi}} \over
{(XY)^{1/2}}}, \
\rho_{nn} = {{(r + \hat{Q})(r + \hat{P})} \over {(XY)^{1/2}}}
e^{{{2(n-2)}\over {n\alpha}}\varphi},
\label{gensol}
\end{eqnarray}
where
\begin{eqnarray}
X &=& r^2 + [(2\beta - \hat{P} + \hat{Q})
{\rm cosh}^2 \delta_2 + \hat{P}]r  + 2\beta \hat{Q}
{\rm cosh}^2 \delta_2,
\nonumber \\
Y &=& r^2 + [(2\beta + \hat{P} - \hat{Q})
{\rm cosh}^2 \delta_1 + \hat{Q}]r  + 2\beta \hat{P}
{\rm cosh}^2 \delta_1,
\nonumber \\
W &=& r^2 + [(2\beta + \hat{P} - \hat{Q}){\rm cosh}^2 \delta_1
+(2\beta - \hat{P} + \hat{Q}){\rm cosh}^2 \delta_2]r +
|P||Q|{\rm cosh} \delta_1 {\rm cosh} \delta_2 {\rm sinh}
\delta_1 {\rm sinh} \delta_2\nonumber\\
& &+2[\beta(2\beta - \hat{P} - \hat{Q}) +\hat{P}\hat{Q}]
{\rm cosh}^2\delta_1 {\rm cosh}^2 \delta_2
+(2\beta - \hat{Q})\hat{P}{\rm cosh}^2 \delta_1 +
(2\beta - \hat{P})\hat{Q}{\rm cosh}^2 \delta_2 ,
\nonumber \\
Z &=& [|P| {\rm sinh} \delta_1 {\rm cosh} \delta_2
+ |Q|{\rm sinh} \delta_2 {\rm cosh} \delta_1 ]r+ |P|\hat{Q}
{\rm sinh} \delta_1 + \hat{P}|Q| {\rm sinh} \delta_2 ,
\label{wxyz}
\end{eqnarray}
with the non-zero electric and magnetic charges and the ADM mass given
by
\begin{eqnarray}
P_{n-1} &=& |P|{\rm cosh} \delta_1 {\rm cosh} \delta_2
+ |Q| {\rm sinh} \delta_1 {\rm sinh} \delta_2 ,\ \
P_n = -(\hat{P} - \hat{Q} +2\beta){\rm cosh} \delta_1 {\rm sinh}
\delta_1 ,
\nonumber \\
Q_{n-1} &=& -(\hat{P} - \hat{Q} - 2\beta){\rm cosh} \delta_2
{\rm sinh} \delta_2 ,\ \
Q_n =|Q|{\rm cosh} \delta_1 {\rm cosh} \delta_2
+ |P|{\rm sinh} \delta_1 {\rm sinh} \delta_2 ,
\nonumber \\
M &=& (2\beta + \hat{P} -\hat{Q}){\rm cosh}^2 \delta_1
+ (2\beta + \hat{Q} - \hat{P}){\rm cosh}^2 \delta_2
+ \hat{P} + \hat{Q} -4\beta ,
\label{genpar}
\end{eqnarray}
where $\hat{P} \equiv \beta + \sqrt{P^2 + \beta^2}$ and
$\hat{Q} \equiv \beta + \sqrt{Q^2 + \beta^2}$.
The requirement of zero Taub-NUT charge relates the
two boost parameters $\delta_{1,2}$ in the following way:
\begin{equation}
|P|{\rm tanh} \delta_2 + |Q|{\rm tanh} \delta_1 = 0.
\label{boostcon}
\end{equation}

The most general solution in this class is finally obtained
by performing $SO(n)/SO(n-2)$ rotations on (\ref{gensol}),
thus providing the remaining $2n-3$ degrees of freedom.

Thus, the general class of solutions is
parameterized in terms of the following $2n+1$ parameters:
the non-extremality parameter $\beta \geq 0$, magnetic $P$ and electric
$Q$ charges of the $U(1)_M \times U(1)_E$ black hole solution,
two $SO(1,1)$ boost parameters $\delta_{1,2}$, which are subject to
the zero Taub-Nut constraint (\ref{boostcon}), and $2n-3$ parameters
of $SO(n)/SO(n-2) \subset SO(n)$ symmetry transformations (\ref{son})
of the four-dimensional action.

One can show that with the zero Taub-NUT constraint (\ref{boostcon})
$M \ge M_{\rm BPS} \equiv |{\vec P}|+|{\vec Q}|$ for $\beta\ge 0$.
For this case $\vec{P} \cdot \vec{Q} \propto \beta$.
Thus the solutions with $\beta=0$ and  other parameters finite
correspond to the supersymmetric solutions discussed in the previous
subsection.  On the other hand for $\delta_{1,2} \to \infty$ and
$|Q|-|P| \to 0$, as $\beta \to 0$, in such a way that
$\beta e^{2|\delta_{1,2}|} \equiv 2|q|$ and $||Q|-|P||
e^{2|\delta_{1,2}|}\equiv 4|\Delta|$ remain finite, one
obtains non-supersymmetric extreme solutions, {\it i.e.},
those with $\beta =0$, however, $\vec{P} \cdot \vec{Q} \neq 0$.

Since the $SO(n)/SO(n-2)\subset SO(n)$ symmetry transformations
with $2n-3$ parameters do not affect the four-dimensional
space-time metric, the four-dimensional properties of the
solution are fully determined by four parameters:
$\beta$, $P$, $Q$ and $\delta_1$ [or $\delta_2$].
Without loss of generality we assume that $|Q| \geq |P|$; for solutions
with $|Q|\leq |P|$, the roles of $\delta_1$ and $\delta_2$ are interchanged.
The Hawking temperature $T_H = \partial_r \lambda |_{r=0}/(2\pi)$
and the entropy  $S=$ (1/4 of the area of the event horizon) are of
the following form:
\begin{equation}
T_H ={{[|Q|^2{\rm cosh}^2 \delta_2
- |P|^2{\rm sinh}^2 \delta_2]^{1/2}} \over {4\pi\left(\hat{P}
\hat{Q}\right)^{1/2}|Q|{\rm cosh}^2 \delta_2}}, \ \ \
S = {{2\pi\beta \left(\hat{P}\hat{Q}\right)^{1/2}
|Q|{\rm cosh}^2 \delta_2} \over {[|Q|^2{\rm cosh}^2 \delta_2 -
|P|^2{\rm sinh}^2 \delta_2]^{1/2}}} .
\label{tempent}
\end{equation}

The thermal properties and the singularity structure of the whole
class of the solutions can be summarized according to the values
of parameters $\delta_2$, $P$ and $\beta$ as \cite{CYALL}:

\begin{itemize}
\item Non-extreme black holes with $\delta_2 \ne 0$ and $P \ne 0$:\\
The global space-time is that of non-extreme Reissner-Nordstr\" om
black holes, {\it i.e.}, the time-like singularity is hidden behind
the inner horizon.  $T_H$ [$S$] is finite, and decreases
[increases] as $\delta_2$ or $\beta$ increases,
approaching zero  [infinity].
\item Non-extreme black holes with $\delta_2 = 0$ or $P = 0$: \\
The singularity structure is that of the Schwarzschield
black holes, {\it i.e.}, the space-like singularity is
hidden behind the (outer) horizon.  $T_H$ [$S$] is finite and decreases
[increases] as $\beta$ increases, approaching zero [infinity].
\item  Supersymmetric extreme black holes, {\it i.e.}, $\delta_2$
finite: \\
For $P\ne 0$, the solution has a null singularity, which becomes
naked  when $P=0$.   $T_H$ [$S$] is finite
and becomes infinite [zero] when $P=0$.
\item  Non-supersymmetric extreme black holes, {\it i.e.},
$|\delta_2| \to \infty$ with ($q$,$\Delta$) finite:\\
The global space-time is that of extreme Reissner-Nordstr\" om
black holes with zero $T_H$ and finite S.
\end{itemize}

\section{Spherically Symmetric Black Holes of Effective
Four-Dimensional $N=4$ Supersymmetric String Vacua}

In the following we shall summarize the results for a
class of BPS saturated solutions \cite{CYHET,CYHETS} and the corresponding
non-extreme configurations parameterized in terms of the fields
of the heterotic string compactified on a six-torus.

The effective field theory of massless bosonic fields for the
heterotic string on a Narain torus \cite{NARAIN} at a generic point
of moduli space is obtained by compactifying the ten-dimensional
$N=1$ Maxwell/Einstein supergravity theory on a six-torus
\cite{SCHWARZ,SEN2}.  The effective four-dimensional action
\footnote{See Refs. \cite{SCHWARZ,SEN2} for notational conventions and
the relationship of four-dimensional fields to the corresponding
ten-dimensional ones.  Also, we are not addressing
$\alpha'$ corrections.}
for massless bosonic fields consists of the graviton $g_{\mu\nu}$, 28
$U(1)$ gauge fields ${\cal A}^i_{\mu} \equiv (A^{(1)\, m}_{\mu},
A^{(2)}_{\mu\, m}, A^{(3)\, I}_{\mu})$, corresponding to the $U(1)$
gauge fields of dimensionally reduced  ten-dimensional
metric (Kaluza-Klein sector), two-form fields, and
Yang-Mills fields, respectively, and 134 scalar fields.
The scalar fields consist of the dilaton $\phi$
(which parameterizes the strength of the string coupling), the axion field
$\Psi$ (which is obtained from the two-form field $B_{\mu\nu}$ through the
duality transformation), and a symmetric $O(6,22)$ matrix $M$ of
132 scalar fields (moduli fields whose vacuum expectation values
parameterize the string vacua).  The matrix $M$ consists of 21 internal
metric $g_{mn}$ components, 15 pseudo-scalar fields $B_{mn}$,
and 96 scalar fields $a^I_m$, which arise from the dimensionally
reduced ten-dimensional metric, two-form field and Yang Mills
fields, respectively.  Here, $(\mu,\nu)=0,\cdots,3$, $(m,n)=1,\cdots, 6$
and $I=1,\cdots,16$.

The four-dimensional effective action is invariant under the $O(6,22,R)$
transformations \cite{SCHWARZ,SEN2}:
\begin{equation}
M \to \Omega M \Omega^T ,\ \ \ {\cal A}^i_{\mu} \to \Omega_{ij}
{\cal A}^j_{\mu}, \ \ \ g_{\mu\nu} \to g_{\mu\nu}, \ \ \ \phi \to \phi .
\label{tdual}
\end{equation}
Here, $\Omega \in O(6,22)$, {\it i.e.}, $\Omega^T L \Omega = L$,
where $L$ is an $O(6,22)$ invariant matrix.
The world-sheet instanton effects break $O(6,22,R)$ invariance
of the effective action down to its discrete subgroup $O(6,22,Z)$
referred to as $T$-duality. $T$-duality is an exact string symmetry to all
orders in string perturbation and is assumed to survive
non-perturbative corrections.

In addition, the equations of motion and Bianchi identities are invariant
under the $SL(2,R)$ transformations \cite{STRWK,SEN2}:
\begin{equation}
S \to S^{\prime}={{aS+b}\over{cS+d}},\ M\to M ,\ g_{\mu\nu}\to g_{\mu\nu},\
{\cal F}^i_{\mu\nu} \to {\cal F}^{\prime\, i}_{\mu\nu} =
(c\Psi + d){\cal F}^i_{\mu\nu} + ce^{-\phi} (ML)_{ij}
\tilde{\cal F}^j_{\mu\nu},
\label{sdual}
\end{equation}
where $S \equiv \Psi + i e^{-\phi}$, $\tilde{\cal F}^{i\,\mu\nu} =
{1\over 2}(\sqrt{-g})^{-1} \varepsilon^{\mu\nu\rho\sigma}
{\cal F}^i_{\rho\sigma}$, and $a,b,c,d \in R$ satisfy $ad-bc=1$.
The space-time instanton effects break $SL(2,R)$ down to $SL(2,Z)$,
referred to as $S$-duality.  $S$-duality, which is non-perturbative
in nature, is conjectured \cite{STRWK} to be an exact symmetry of
$N=4$ supersymmetric string vacua. It  relates  strongly coupled
vacua to those  of the weakly coupled ones.

The allowed discrete magnetic ${\vec P}$ and electric ${\vec Q}$
charges are determined \cite{SEN2} by  $T$- and $S$-duality constraints
 of  toroidally compactified heterotic
string and by the
Dirac-Schwinger-Zwanzinger-Witten (DSZW) quantization condition
\cite{DSZ,WITTENIII}; both of the ``lattice charge vectors'' \cite{SEN2},
$\vec{\beta}\equiv L \vec{P}$ and $\vec{\alpha}\equiv
{\rm e}^{-\phi_{\infty}}M^{-1}_{\infty}\vec{Q}-\Psi_{\infty} \vec{\beta}$,
belong to an even, self-dual Lorentzian lattice $\Lambda$ with the
signature $(6,22)$.  Here $\phi_{\infty}$, $\Psi_{\infty}$ and
$M_{\infty}$ correspond to the asymptotic values of the dilaton,
axion and moduli fields, respectively.  Note, that the lattice
charge vectors $\vec{\alpha}$ and $\vec{\beta}$ tranform covariantly
\cite{SEN2} under $T$- and $S$-duality transformations.

Within this effective theory, we shall present explicit results for a
class of spherically symmetric BPS saturated configurations
\cite{CYHET} as well as a class of their non-extreme counterparts.
The Killing spinor equations for the BPS saturated states are
obtained by setting to zero the ten-dimensional supersymmetry
transformations for the gravitino, dilatino and 16 gaugini,
now expressed in terms of the four-dimensional fields.
The spectrum of BPS saturated, static, spherically symmetric
configurations is both $O(6,22,Z)$ and $SL(2,Z)$ invariant.
Namely, the Killing spinors $\varepsilon$ are invariant under
$T$-duality and transform covariantly under $S$-duality.
Therefore one can generate new BPS saturated solutions as well as
 non-extreme solutions by imposing $S$- and $T$-duality
transformations on a particular  ``generating''  solution.

For the purpose of obtaining a general set of solutions with
an arbitrary choice of $M_\infty$ and $S_{\infty}=\Psi_{\infty}+
i{\rm e}^{-\phi_{\infty}}$, one can use the following procedure:
\begin{itemize}
\item
First one performs \cite{SEN2,HTII} $O(6,22,R)$ transformations
of the form $M_{\infty} \to \hat{M}_{\infty} = \hat{\Omega}
M_{\infty}\hat{\Omega}^T$ and $\Lambda \to \hat{\Lambda}=
L\hat{\Omega} L\Lambda$ ($\hat{\Omega} \in O(6,22,R)$).
This procedure allows one to bring an arbitrary asymptotic value of
the moduli fields $M$ to the identity matrix, {\it i.e.},
$\hat{M}_{\infty} = I_{28}$, while the electric and magnetic
lattice charge vectors live in the new lattice $\hat{\Lambda}$.
Note, that the transformation $\Omega$ that gives $\hat{\Lambda}$
is determined only up to the $O(6,22,Z)$ automorphisms of the lattice
$\Lambda$.  A subset of $O(6,22,Z)$ transformations that preserves the
asymptotic value $\hat{M}_{\infty}=I_{28}$ is $SO(6,Z)\times SO(22,Z)$.
The latter subset of $T$-duality transformations generates
$O(6,22,Z)$ orbits of solutions with the same $\hat{M}_{\infty}=I_{28}$.
\item
Secondly one can use $SL(2,R)$ transformations to bring $S_{\infty}
\to \breve{S}_\infty=i$ and correspondingly transform the lattice
charge vectors living in $\hat{\Lambda}$ into those living in a new
lattice $\breve{\Lambda}$.  A subset of $S$-duality transformations
that preserves the asymptotic value $\breve{S}_{\infty}=i$ are
$SO(2,Z)$ transformations which generate $SL(2,Z)$ orbits of
solutions with the same $\breve{S}_{\infty}=i$.
\item
Finally, in order to obtain a set of solutions with arbitrary
asymptotic values of $M$ and $S$, one has to undo the above
$O(6,22,R)$ and $SL(2,R)$ transformations.
\end{itemize}

Using the procedures described above, we shall now present a class
of dyonic BPS saturated states (and their  non-extreme
counterparts), which correspond to $O(6,22,Z)$ and $SL(2,Z)$ orbits
of the most general dyonic solution with zero axion.
When the axion field is turned off, the Killing spinor equations
ensure that the BPS saturated states with 28 electric $\vec{Q}$
and 28 magnetic $\vec{P}$ charges are subject to two (orthogonality)
constraints \cite{CYHET}:
\begin{equation}
\vec{P}^T{\cal M}_{\pm}\vec{Q}=0\ \ \ \
({\cal M}_{\pm} \equiv (LM_{\infty}L\pm L)).
\label{gencon}
\end{equation}
The $S$-duality transformations provide  one more parameter for the
charge degrees of freedom along with the non-zero axion field.
In this case $28$ electric and $28$ magnetic charges are subject
to the following one constraint \cite{CYHET}:
\begin{equation}
\vec{P}^T{\cal M}_{-}\vec{Q}\,[\vec{Q}^T{\cal M}_{+}\vec{Q} -
\vec{P}^T{\cal M}_{+}\vec{P}] -\vec{P}^T{\cal M}_{+}\vec{Q}
[\vec{Q}^T{\cal M}_{-}\vec{Q} - \vec{P}^T{\cal M}_{-}\vec{P}] = 0.
\label{fgencon}
\end{equation}
The general class of solutions subject to the charge constraint
(\ref{fgencon}) is obtained following the procedure described above:

\noindent{({\it i}) First, on a solution with chosen asymptotic values
$M_{\infty}$ and $S_{\infty}$ one performs the above mentioned $SL(2,R)$
and $O(6,22,R)$ transformations, rendering $S_\infty\to \breve{S}_\infty
=i$ and $M_{\infty}\to \hat{M}_{\infty}=I_{28}$, respectively, along with
the corresponding transformations of the charge lattices.}

\noindent{({\it ii}) The generating solution for a general class of
solutions with the charge constraint (\ref{gencon}) (and with $\hat M_\infty
=I_{28}$ and $\breve S_\infty=i$) turns out to  correspond to the
$\breve{U}(1)^{(1)}_{m,\,M} \times \breve{U}(1)^{(1)}_{n,\,E} \times
\breve{U}(1)^{(2)}_{m,\,M} \times \breve{U}(1)^{(2)}_{n,\,E}$
configuration ($1 \leq m \neq n \leq 6$) \cite{CYHET}.  Namely this
configuration is parameterized by two magnetic and two electric charges
(with the corresponding lattice charge vectors living in lattice
$\breve{\Lambda}$), which arise from different $U(1)$ groups; the
two magnetic [electric] charges arise from the Kaluza-Klein sector
gauge field $\breve{A}^{(1)\, m}_{\phi}$ [$\breve{A}^{(1)\, n}_{t}$]
and the corresponding two-form $U(1)$ field
$\breve{A}^{(2)}_{\phi\, m}$ [$\breve{A}^{(2)}_{t\, n}$].  Without loss
of generality we choose the non-zero charges to be $\breve{P}^{(1)}_1,
\breve{P}^{(2)}_1, \breve{Q}^{(1)}_2, \breve{Q}^{(2)}_2$.}

\noindent{({\it iii}) $[SO(6,Z)\times SO(22,Z)]/[SO(4,Z)\times SO(20,Z)]$
transformations, {\it i.e.}, a subset of $O(6,22,Z)$ transformations
preserving the asymptotic value $\hat{M}_{\infty}=I_{28}$, on the
generating solutions provide 50 additional parameters specifying
the $O(6,22,Z)$ orbits with the general charges subject to two
orthogonality constraints (\ref{gencon}).}

\noindent{({\it iv}) In addition $SO(2,Z)$ transformations,
{\it i.e.}, a subset of $SL(2,Z)$ transformations preserving
$\breve S_{\infty}$, provide  one more parameter specifying
$SL(2,Z)$ orbits consistent with the constraint (\ref{fgencon}).}

\noindent{({\it v}) Finally one has to undo the $SL(2,R)$ and
$O(6,22,R)$ transformations to obtain configurations with chosen
asymptotic values $M_\infty$ and $S_\infty$.}

Note that the set of transformations used in the above procedure
{\it does not affect the four-dimensional space-time} and thus all
the solutions in the class have the same four-dimensional
space-time structure.

The explicit form for the static, spherically symmetric generating
solution is \cite{CYHET}:
\begin{eqnarray}
\lambda &=& r^2/[(r-\eta_P \breve{P}^{(1)}_1)
(r-\eta_P \breve{P}^{(2)}_1)
(r- \eta_Q \breve{Q}^{(1)}_2)
(r-\eta_Q \breve{Q}^{(2)}_2)]^{1\over 2},
\nonumber\\
R &=& [(r-\eta_P \breve{P}^{(1)}_1)
(r - \eta_P \breve{P}^{(2)}_1)
(r - \eta_Q \breve{Q}^{(1)}_2)
(r-\eta_Q \breve{Q}^{(2)}_2)]^{1\over 2},
\nonumber\\
e^{\phi}&=&\left [{(r-\eta_P \breve{P}^{(1)}_1)
(r- \eta_P \breve{P}^{(2)}_1)} \over
{(r- \eta_Q \breve{Q}^{(1)}_2)
(r- \eta_Q \breve{Q}^{(2)}_2)}\right]^{1\over 2},\ \ \Psi=0,
\nonumber\\
g_{11}&=&{{r- \eta_P \breve{P}^{(2)}_1} \over
{r-\eta_P \breve{P}^{(1)}_1}}, \
g_{22}={{r- \eta_Q \breve{Q}^{(1)}_2} \over
{r- \eta_Q \breve{Q}^{(2)}_2}},\
g_{mm}=1\  \ (m \neq 1,2) ,\nonumber\\
 g_{mn}&=&B_{mn}=0\ \ (m\ne n),\ \ \ a^I_m=0.
\label{hetgensol}
\end{eqnarray}
Here the radial coordinate is chosen so that the horizon is at $r=0$.
$\eta_{P,Q}=\pm$ correspond to parameters appearing in the Killing
spinor constraints.  Namely the upper $\varepsilon_{u}$ and lower
$\varepsilon_{\ell}$ two-component Killing spinors are subject to
the constraints \cite{CYHET}: $\Gamma^1 \varepsilon_{u,\ell}=
i\eta_P\varepsilon_{\ell,u}$ if $\breve{P}^{(1)}_1 \ne 0 $ and/or
$\breve{P}^{(2)}_1 \ne 0$, and $\Gamma^2\varepsilon_{u,\ell}=\mp
\eta_Q \varepsilon_{\ell,u}$ if $\breve{Q}^{(1)}_2 \ne 0 $ and/or
$\breve{Q}^{(2)}_2 \ne 0$.  Thus non-zero magnetic and electric
charges each break ${1\over 2}$ of the remaining supersymmetries;
purely electric [or magnetic] and dyonic configurations preserve
${1\over 2}$ and ${1\over 4}$ of $N=4$ supersymmetry, respectively.
The first and the second sets of configurations fall into vector and
highest spin ${3\over 2}$ supermultiplets \cite{KALL}, respectively.

The requirement that the ADM mass of the above configurations saturates
the Bogomol'nyi bound restricts the choice of parameters $\eta_{P,Q}$ to
be such that $\eta_P\, {\rm sign}(\breve{P}^{(1)}_1+\breve{P}^{(2)}_1)=-1$
and $\eta_Q\, {\rm sign}(\breve{Q}^{(1)}_2+\breve{Q}^{(2)}_2)=-1$, thus
yielding the positive semidefinite ADM mass of the following form
\footnote{General BPS saturated states have the following
$O(6,22,Z)$ and $SL(2,Z)$ invariant form of the ADM mass \cite{CYHET}:
$M^2_{\rm BPS} = e^{-\phi_\infty}\{{\vec P}^T {\cal M}_{+} {\vec P} +
{\vec Q}^T {\cal M}_{+} {\vec Q}+ 2[({\vec P}^T{\cal M}_{+}{\vec P})
({\vec Q}^T{\cal M}_{+}{\vec Q})-({\vec P}^T{\cal M}_{+}
{\vec Q})^2]^{1\over 2}\} $, which is a generalization of the one for
BPS states preserving ${1\over 2}$ of the original supersymmetry.}:
\begin{equation}
M_{\rm BPS} = |\breve{P}^{(1)}_1+\breve{P}^{(2)}_1|
+|\breve{Q}^{(1)}_2+\breve{Q}^{(2)}_2|.
\label{ADMmass}
\end{equation}

One can also obtain the non-extreme solutions, parameterized by
the above four non-zero charges and the non-extremality parameter
$\beta$, by solving the Einstein field equations and
Euler-Lagrange equations, which yield the following result:
\begin{eqnarray}
\lambda &=& r(r+2\beta)/[(r+\breve{P}^{(1)\,\prime}_1)
(r+\breve{P}^{(2)\,\prime}_1)(r+\breve{Q}^{(1)\,\prime}_2)
(r+\breve{Q}^{(2)\,\prime}_2)]^{1\over 2},
\nonumber\\
R(r) &=& [(r+\breve{P}^{(1)\,\prime}_1)(r+\breve{P}^{(2)\,\prime}_1)
(r+\breve{Q}^{(1)\,\prime}_2)(r+\breve{Q}^{(2)\,\prime}_2)]^{1\over 2},
\nonumber\\
e^{\phi} &=& \left [ {{(r+\breve{P}^{(1)\,\prime}_1)
(r+\breve{P}^{(2)\,\prime}_1)} \over {(r+\breve{Q}^{(1)\,\prime}_2)
(r+\breve{Q}^{(2)\,\prime}_2)}}\right ]^{1\over 2},\ \Psi=0,
\nonumber\\
g_{11} &=& {{r+\breve{P}^{(2)\,\prime}_1}\over
{r+\breve{P}^{(1)\,\prime}_1}}, \ \
g_{22} = {{r+\breve{Q}^{(1)\,\prime}_2}\over
{r+\breve{Q}^{(2)\,\prime}_2}},\ \
g_{mm} = 1\ \ (m \neq 1,2),\nonumber\\
 g_{mn}&=&B_{mn}=0\ \  (m\ne n),\ \ \ a^I_m=0,
\label{hetnonex}
\end{eqnarray}
with the ADM mass:
\begin{equation}
M_{\text{non-ext}}= \breve{P}^{(1)\,\prime}_1 + \breve{P}^{(2)\,\prime}_1 +
\breve{Q}^{(1)\,\prime}_2 + \breve{Q}^{(2)\,\prime}_2-4\beta .
\label{nonexmass}
\end{equation}
Here $\breve{P}^{(1)\,\prime}_1 \equiv \beta \pm \sqrt{(\breve{P}^{(1)}_1)^2
+ \beta^2}$, {\it etc}. and $\beta$ parameterizes the deviation from the
corresponding supersymmetric solutions.

The signs ($\pm$) in the expressions for $\breve{P}^{(1)\,\prime}_1$,
{\it etc.} should be chosen so that in the limit  $\beta\to 0$,
$M_{\text{non-ext}}\to M_{\rm BPS}$.  Note that $O(6,22,Z)$ and
$SL(2,Z)$ orbits of the non-extreme solutions are subject to the
charge constraint (\ref{fgencon}) and thus constitute only a subset
of a general class of non-extreme configurations.
The full set of non-extreme solutions should be obtained by performing
a subset of $O(8,24,Z)$ transformations, {\it i.e.}, symmetry
transformations of the effective three-dimensional action \cite{THREE},
on (\ref{hetnonex}).

In the following two subsections we shall address the four-dimensional
space-time structure of these solutions.

\subsection{Regular Dyonic Solutions}

For a black hole solution, {\it i.e.}, a spherically symmetric
configuration with the regular horizon, one has to choose the
relative signs of the two electric and two magnetic charges of
the BPS saturated generating solutions (\ref{hetgensol}) to be the
same \cite{CYHET}, so that the space-time singularity (the point
at which $R(r)$ vanishes) lies inside or on the horizon (the point
at which $\lambda(r)$ vanishes).  In this case the  non-extreme
solutions are given by (\ref{hetnonex}) with the {\it positive } signs
in the expressions for $\breve{P}^{(1)\,\prime}_1$, {\it etc.}
and have the ADM mass:
\begin{equation}
M_{\text{non-ext}}=\sqrt{(\breve{P}^{(1)}_1)^2+\beta^2}+
\sqrt{(\breve{P}^{(2)}_1)^2 +\beta^2}+
\sqrt{(\breve{Q}^{(1)}_2)^2+\beta^2}+\sqrt{(\breve{Q}^{(2)}_2)^2
+\beta^2},
\label{regnonmass}
\end{equation}
which is always compatible with the Bogomol'nyi bound:
\begin{equation}
M_{\rm BPS}=|\breve{P}^{(1)}_1|+|\breve{P}^{(2)}_1|+|\breve{Q}^{(1)}_2|
+|\breve{Q}^{(2)}_2|.
\label{regexmass}
\end{equation}
These solutions always have nonzero mass
\footnote{Interestingly one can draw parallels between the relation of
regular dyonic BPS saturated states to their non-extreme
counterparts and that of Type-II supergravity walls \cite{CGNPB,EXWALL}
to their non-extreme counterparts \cite{PRL,WALL}.  Type-II supergravity
walls \cite{EXWALL} are planar configurations in $N=1$ supergravity
theory, interpolating between two isolated supersymmetric
anti-deSitter vacua, whose ADM mass density (in the thin wall
approximation) is determined to be $\kappa\sigma_{\text{ext}}=
2(\alpha_1 + \alpha_2)$, where $\Lambda_i \equiv -3\alpha^2_i$ is
the cosmological constant on each side of the wall.  The corresponding
set of non-extreme domain wall solutions \cite{PRL,WALL}
corresponds to spherically symmetric bubbles with two insides whose
ADM mass density  $\kappa\sigma_{\text{non-ext}} =
2(\sqrt{\alpha_1^2+\beta^2} + \sqrt{\alpha_2^2+\beta^2})$ is larger
than that of the Type-II supergravity wall.  Here $\beta$
parameterizes, analogously as in the case of the non-extreme black
holes, a deviation from the extreme limit.}.

The singularity structure of  and thremal properties of regular
solutions  can be summarized in the following way \cite{CYHET}:
\begin{itemize}
\item
The case with {\it all the four charges non-zero} corresponds to
black holes with two horizons at $r=0,-2\beta$ and a time-like
singularity hidden behind the inner horizon, {\it i.e.}, the
global space-time is that of the Reissner-Nordstr\" om black
holes.  The Hawking temperature is $T_H=\beta/(\pi
\sqrt{\breve{P}^{(1)\,\prime}_1\breve{P}^{(2)\,\prime}_1
\breve{Q}^{(1)\,\prime}_2 \breve{Q}^{(2)\,\prime}_2})$ and the entropy
is finite $S=\pi\breve{P}^{(1)\,\prime}_1 \breve{P}^{(2)\,\prime}_1
\breve{Q}^{(1)\,\prime}_2 \breve{Q}^{(2)\,\prime}_2$. As $\beta \to 0$
the space-time is that of extreme Reissner-Nordstr\" om black holes.
\item
The case with {\it three nonzero charges}, say, $\breve{P}^{(1)}_1=0$,
corresponds to solutions with a space-like singularity located at the inner
horizon ($r=-2\beta$), $T_H=\beta^{1\over 2}/(\pi
\sqrt{2\breve{P}^{(2)\,\prime}_1 \breve{Q}^{(1)\,\prime}_2
\breve{Q}^{(2)\,\prime}_2})$ and $S=2\pi\beta \breve{P}^{(2)\,\prime}_1
\breve{Q}^{(1)\,\prime}_2 \breve{Q}^{(2)\,\prime}_2$.  As $\beta \to 0$
the singularity coincides with the horizon at $r=0$.
\item
The case with {\it two charges nonzero}, say, $\breve{P}^{(1)}_1
\ne 0 \ne \breve{P}^{(2)}_1$,
corresponds to solutions with a space-like singularity  at $r=-2\beta$,
$T_H=1/(2\pi\sqrt{\breve{P}^{(1)\,\prime}_1 \breve{P}^{(2)\,\prime}_1})$
and $S=4\pi\beta^2 \breve{P}^{(1)\,\prime}_1 \breve{P}^{(2)\,\prime}_1$.
As $\beta \to 0$ the singularity coincides with the horizon at $r=0$.
\item
The case with {\it one nonzero charge}, say, $\breve{P}^{(1)}_1 \neq 0$,
corresponds to black holes with a space-like singularity at $r=-2\beta$,
$T_H=1/(2\pi\sqrt{2\beta \breve{P}^{(1)\,\prime}_1})$ and $S=8\pi\beta^3
\breve{P}^{(1)\,\prime}_1$.  As $\beta \to 0$ the singularity
becomes naked.
\end{itemize}

\subsection{Singular BPS Saturated Solutions}

When the relative signs for the two magnetic and/or two electric
charges are opposite \cite{CYHETS} the BPS saturated generating solutions
(\ref{hetgensol}) are always singular, {\it i.e.}, the singularity
takes place at $r_{\rm sing}>0$
\footnote{Such purely electrically charged configurations are related
to massless black holes, recently found by Behrndt \cite{BEHRNDT2},
which were obtained by dimensionally reducing supersymmetric gravitational
waves of the effective ten-dimensional heterotic string theory.
Generalizations to the corresponding multi-black hole solutions
and the corresponding exact (in $\alpha'$ expansion) magnetic
solutions were given by Kallosh \cite{KALL1}, and by Kallosh and Linde
\cite{KALLIND}, respectively.  In the latter work the physical
properties of such configurations were further addressed;
they repel massive particles.}.
In this case the  non-extreme solutions are given by (\ref{hetnonex})
with the negative sign in either $\breve{P}^{(1)\,\prime}_1$
[and/or $\breve{Q}^{(1)\,\prime}_2$]  or $\breve{P}^{(2)\,\prime}_1$
[and/or $\breve{Q}^{(2)\,\prime}_2$], in such a way that their ADM
mass (\ref{nonexmass}) reduces to that of the BPS saturated solutions
(\ref{ADMmass}) as $\beta \to 0$.  In particular, when the relative
signs of the two magnetic and two electric charges are opposite,
the ADM mass for non-extreme solutions:
\begin{equation}
M_{\text{ultra-ext}} = \left |\sqrt{(\breve{P}^{(1)}_1)^2+\beta^2} -
\sqrt{(\breve{P}^{(2)}_1)^2 +\beta^2}\right | +
\left |\sqrt{(\breve{Q}^{(1)}_2)^2+\beta^2}
- \sqrt{(\breve{Q}^{(2)}_2)^2 +\beta^2}\right |
\label{singnonmass}
\end{equation}
is always {\it smaller} than the ADM mass for the corresponding
BPS saturated states
\footnote{One can also draw parallels between the relation of the
singular dyonic BPS saturated states to their non-extreme
counterparts and that of Type-III supergravity  walls
\cite{CGNPB,EXWALL} to their ultra-extreme counterparts
\cite{PRL,WALL}.  Type-III supergravity walls in $N=1$ supergravity
correspond \cite{CGNPB,EXWALL} to planar configurations,
interpolating between two specific isolated supersymmetric
anti-deSitter vacua with cosmological constants $\Lambda_i\equiv
-3 \alpha^2_i$, whose ADM mass density is given by $\kappa
\sigma_{\text{ext}}=2(\alpha_1 - \alpha_2)$ $(\alpha_1 > \alpha_2)$.
The set of non-extreme solutions corresponds to bubbles of the
false vacuum decay whose ADM mass density $\kappa
\sigma_{\text{ultra-ext}}=2(\sqrt{\alpha^2_1 + \beta^2} -
\sqrt{\alpha^2_2 + \beta^2})$ is always {\it smaller} than that of
Type-III supergravity walls.}:
\begin{equation}
M_{\rm BPS}=\left ||\breve{P}^{(1)}_1|-|\breve{P}^{(2)}_1|\right| +
\left ||\breve{Q}^{(1)}_2| - |\breve{Q}^{(2)}_2|\right |.
\label{singexmass}
\end{equation}
Thus these non-extreme solutions are not in the spectrum;
singular BPS solutions with the relative signs for both the two
electric and two magnetic charges opposite {\it do not have
non-extreme counterparts} compatible with the corresponding
Bogomol'nyi bound.

These BPS saturated solutions are singular with a naked singularity
at $r=r_{\rm sing}\equiv{\rm max}\{{\rm min}[|\breve{P}^{(1)}_1|,
|\breve{P}^{(2)}_1|],{\rm min}[|\breve{Q}^{(1)}_2|,
|\breve{Q}^{(2)}_2|]\} >0$.  They repel massive particles,
just like in the special case of purely electric [or purely magnetic]
solutions\cite{KALLIND}, however, massless particles with zero
angular momentum reach a naked singularity in a finite proper time.
In addition these configurations become massless \cite{CYHETS} when
the magnitudes of the two magnetic and the two electric charges are
equal, {\it i.e.}, when $|\breve{P}^{(1)}_1|= |\breve{P}^{(2)}_1|$
and $|\breve{Q}^{(1) }_2|= |\breve{Q}^{(2) }_2|$.

There are also hybrid singular solutions with the opposite relative
signs for one type of charges, say, magnetic ones with
$|\breve P^{(1)}_1|>|\breve{P}^{(2)}_1|$, and the same relative
signs for the other type of charges, {\it i.e.}, electric ones.
These solutions are singular with a singularity, say, at
$r_{\rm sing}=\sqrt{(\breve P^{(2)}_1)^2 + \beta^2} - \beta>0$, however,
the non-extreme solutions with $M_{\text{non-ext}}\ge M_{\rm BPS}$
are in the spectrum, provided $\sqrt{(\breve P^{(2)}_1)^2+\beta^2}
\left(1/\sqrt{(\breve P^{(1)}_1)^2+\beta^2}+1/ \sqrt{(\breve Q^{(1)}_2)^2+
\beta^2}+1/\sqrt{(\breve Q^{(2)}_2)^2+\beta^2})\right) \ge 0$. Note that
this set of solutions always has non-zero ADM mass.

\subsection{Implications of Massless BPS Saturated States for $N=4$
String Vacua}

By exploring the ADM mass formula for the BPS saturated states,
preserving $1\over 2$ of $N=4$ supersymmetry, Hull and Townsend
\cite{Hull,HTII} show that massless BPS saturated states can occur at
certain points of moduli space of four-dimensional $N=4$ effective
superstring vacua, parameterized in terms of fields and allowed
charges of toroidally compactified heterotic string.  Such massless
BPS states contribute to a phenomenon \cite{Hull,HTII} which
is a generalization of the Halpern-Frenkel-Ka\v c (HFK) mechanism;
namely at special points of moduli space along with the perturbative
electrically charged massless string states, which enhance the gauge
symmetry of the perturbative string states to the non-Abelian one,
there are massless BPS saturated magnetic monopoles and a tower of
$SL(2,Z)$ related BPS saturated dyons, which may contribute to a new
phase of the enhanced non-Abelian gauge symmetry.

Singular massless BPS saturated states, whose generating solutions
are  purely magnetic [or purely electric] configurations with magnetic
[or electric] charges $\breve P^{(1),(2)}_1$ [or $\breve Q^{(1),(2)}_2$]
opposite in relative signs and equal in magnitude, provide an
explicit realization of a class of  massless BPS saturated states
contributing the generalized HFK mechanism.  In addition there are
also massless dyonic solutions whose generating solutions have both
the two  magnetic $\breve P^{(1),(2)}_1$ {\it and} the two electric
$\breve Q^{(1),(2)}_2$ charges opposite in signs and equal in magnitudes.
Since such states preserve $1\over 4$ of $N=4$ supersymmetry they belong
to massless highest spin $3\over 2$ supermultiplets and may allow for an
appearance of additional massless gravitini, thus providing hints of a
new phase with enhanced local supersymmetry \cite{CYHETS}.

For the purpose of illustrating the enhancement of symmetries at
particular points of moduli space \cite{CYHETS}, we will choose a set
of solutions with a particularly simple, however, non-trivial choice
for the asymptotic values of the scalar fields ($M$ and $S$) and the
charge configuration
\footnote{Solutions with more general charge configurations and
asymptotic values of scalar fields allow for different enhancements
of symmetries at different points of moduli space.};
$M_{\infty}$ is diagonal, {\it i.e.}, from moduli only the
asymptotic values of the diagonal components of the internal metric,
$g_{mm}$, are turned on, and $S_{\infty}$ is purely imaginary,
{\it i.e.}, the asymptotic value of the axion is set to zero.
In this case, the undoing of the corresponding $O(6,22,R)$ and
$SL(2,R)$ transformations on the generating solutions (\ref{hetgensol})
and (\ref{hetnonex}) yields solutions whose ADM mass is now
expressed in terms of the asymptotic values $M_{\infty}$ and
$S_\infty$, and charge lattice vectors living in $\Lambda$.
The ADM mass is then of the form:
\begin{equation}
M_{\rm BPS} = e^{-{\phi_{\infty} \over 2}}|g^{1\over 2}_{11\,\infty}
\beta^{({2})}_1 + g^{-{1\over 2}}_{11\,\infty}\beta^{(1)}_1| +
e^{{\phi_\infty}\over 2}|g^{{1\over 2}}_{22\,\infty}\alpha^{(1)}_2+
g^{-{1\over 2}}_{22\,\infty}\alpha^{(2)}_2|.
\label{BPSmass}
\end{equation}
The allowed magnetic and electric charge lattice vectors, living in
a self-dual even Lorentzian lattice $\Lambda$, that give rise
to massless black holes are, say, $\beta_1^{(1)}=-\beta_1^{(2)}=\pm 1$
and $\alpha^{(1)}_2=-\alpha^{(2)}_2=\pm 1$.

The $SL(2,Z)$ orbits of the generating solution provide a tower of
dyonic solutions with the non-zero axion field and the same dependence
on the moduli fields $M$, and thus  are massless at the same points of
the moduli space as the   $O(6,22,R)$ and $SL(2,R)$ undone generating
solution.  On the other hand the $O(6,22,Z)$ orbits of the generating
solution provide solutions with a dependence on transformed moduli
fields and thus  become massless at different points of the moduli points.

We first consider the case of BPS saturated states that preserve
$1\over 2$ of supersymmetries.  These are purely, say, electrically
charged states with {\it two} possible charge lattice vectors:
$\alpha^{(1)}_2=-\alpha^{(2)}_2=\pm 1$. Such states become massless at
the self-dual point of a ``one-torus'', {\it i.e.}, when
$g_{22\,\infty}=1$, and form, together with the $U(1)^{(1)+(2)}$
gauge field $(A^{(1)}_{\mu\,2}+A^{(2)}_{\mu\,2})/\sqrt 2$, the adjoint
representation of the non-Abelian $SU(2)^{(1)+(2)}_2$ gauge group, thus
enhancing the gauge symmetry from $U(1)^{(1)}_2 \times U(1)^{(2)}_2$ to
$U(1)^{(1)-(2)}_2 \times SU(2)^{(1)+(2)}_2$ at this point of the moduli
space \cite{CYHETS}.  There is also an infinite tower of massless dyonic
states, including purely magnetic ones, which correspond to $SL(2,Z)$
orbits of the generating solutions, and thus may contribute to a new
phase of enhanced gauge symmetry.

On the other hand for massless BPS saturated states that preserve
${1\over 4}$ of supersymmetries, the local supersymmetry is enhanced
since the corresponding supermultiplets contain the gravitino
as well as the $U(1)$ gauge field.  The possible charge configurations
in the charge lattice that could give rise to the massless states are
$\beta_1^{(1)}=-\beta_1^{(2)}=\pm 1$ and $\alpha^{(1)}_2=-\alpha^{(2)}_2=
\pm 1$.  These states become massless at the self-dual point of the
corresponding two-torus, {\it i.e.}, when $g_{11\,\infty}=g_{22\,\infty}=1$.
Since each of these additional massless states belongs to the highest
spin $3\over 2$ supermultiplete, the local supersymmetry is enhanced
\cite{CYHETS} from $N=4$ to $N=8$.  As in the previous case, these
four massless states combine with the $U(1)^{(1)+(2)}_1$ gauge field
$(A^{(1)}_{\mu\,1} +A^{(2)}_{\mu\,1})/\sqrt 2$ and the
$U(1)^{(1)+(2)}_2$ gauge field $(A^{(1)}_{\mu\,2} +A^{(2)}_{\mu\,2})
/\sqrt 2$ to form the adjoint representation of the non-Abelian
$SU(2)^{(1)+(2)}_1 \times SU(2)^{(1)+(2)}_2$ gauge group, thus enhancing
the gauge symmetry from $U(1)^{(1)}_1 \times U(1)^{(2)}_1 \times
U(1)^{(1)}_2 \times U(1)^{(2)}_2$ to $U(1)^{(1)-(2)}_1\times
U(1)^{(1)-(2)}_2\times SU(2)^{(1)+(2)}_1\times SU(2)^{(1)+(2)}_2$
at this  point of moduli space \cite{CYHETS}.  Additionally there is
an infinite tower of dyonic massless states that are related through
$SL(2,Z)$ transformations.  The occurrence of these new types of
states may indicate a transition to a new phase of superstring
vacua.

\section{Conclusions}

We have discussed a general set of BPS saturated solutions and
their non-extreme counterparts which arise in effective supergravity
theories compactified down to four dimensions on manifolds with Abelian
isometries.  We concentrated on a general class of solutions of the
effective $N=4$ superstring vacua, parameterized in terms of fields
of the effective heterotic string theory compactified on a six-torus.
Such a program was completed within the Kaluza-Klein sector
of the ($4+n$)-dimensional (minimally extended) supergravities
compactified on $n$-tori, {\it i.e.}, by obtaining explicit results
for all the four-dimensional static, spherically symmetric BPS
saturated states \cite{SUPER} as well as all of their non-extreme
solutions \cite{CYALL}. Within four-dimensional effective $N=4$
supersymmetric string vacua we presented a class \cite{CYHET,CYHETS} of
BPS saturated states as well as a class of their non-extreme
counterparts which correspond to $O(6,22,Z)$ and $SL(2,Z)$
orbits of dyonic configurations with zero axion; those are
configurations whose 28 electric and 28 magnetic charges are
subject to one constraint (\ref{fgencon}).

The generating solution is parameterized by two electric and
two magnetic charges and in the non-extreme case additionally
by the non-extremality parameter $\beta$.  The BPS states whose
generating solutions are purely electrically charged [or purely
magnetically charged] and dyonic preserve $1\over 2$ and
$1\over 4$ of $N=4$ supersymmetry, respectively.  These solutions
fall into different classes depending on the relative signs of the
two magnetic and two electric charges of the generating solution.
When the relative signs of both two sets of these charges are the
same \cite{CYHET} solutions are regular (with a horizon in four
dimensions) accompanied by the non-extreme counterparts and always
have the ADM mass non-zero.  On the other hand when the relative
sign of at least one of the two sets of charges of the generating
solution is opposite \cite{CYHETS} solutions are singular
(they have a naked singularity).  In the case when both sets of
charges have opposite relative sings the singular solutions are
unaccompanied by the  non-extreme counterparts (whose ADM masses
are compatible with the Bogomol'nyi bound) and have zero mass when
the magnitudes of the two magnetic and two electric charges are
equal \cite{CYHETS}.

Purely electrically charged BPS saturated states have the same mass
spectrum and charge assignments as a subset of perturbative string
excitations, and should probably be identified with each other
\cite{DUFFR}.  On the other hand the magnetically charged and dyonic
BPS saturated states should be viewed as non-perturbative states of
string vacua.  Massless BPS states with purely magnetically charged
generating solutions (along with a tower of $SL(2,Z)$ orbits) may
contribute to the generalized HFK-mechanism \cite{Hull,HTII}, while
massless BPS states with dyonic genetating  solutions may contribute
to an enhancement of local supersymmetry \cite{CYHETS} at such points
of moduli space.   We would, however, like to caution that at these
points of moduli space the effective theory approach breaks down
due to the appearance of an infinite tower of new massless modes.
In addition since we have studied only classical configurations within
an effective theory, quantum corrections may qualitatively alter
the features of such massless states.  Thus full physical
implications of such massless states await further investigation.

The heterotic string compactified on $T^4$ is conjectured
\cite{DUFFSS,WITTENII,HTI,STRDUAL} to be equivalent to the
type IIA string compactified on $K3$ surface.  Since the
four-dimensional effective actions of these two theories further
compactified on $T^2$ are related through a field redefinition,
one can address  the  BPS saturated solutions in the type IIA
string on $T^2\times K3$.  Whenever the gauge symmetry of the
heterotic string is enhanced to a non-Abelian group at particular
points of moduli space, $K3$ surface of the type IIA string theory
develops quotient singularities (and thus is an orbifold as far as
target space geometry is concerned) with certain homology two-cycles
of $K3$ collapsing to zero area, giving rise to massless BPS saturated
states  \cite{WITTENII,HTII}.  In the case of the type IIA string on
$K3$ surface the point in the moduli space representing the conformal
field theory orbifold and the point representing the theory with the
enhanced gauge symmetry do not coincide, indicating that the conformal
field theory is ill-behaved \cite{K3}, {\it i.e.}, perturbative string
theory does not describe the full string dynamics.

Since the BPS saturated states presented here are derived within the
effective field theory compactified from a higher-dimensional
theory, they should  presumably have origins as p-brain solutions
in higher dimensions \cite{PBRANE}; the case of magnetically
charged BPS states in the dual type IIA string on $K3\times T^2$
was studied in \cite{HTII}.  Also, four-dimensional dilatonic
black holes can be obtained from dimensionally reduced
Brinkmann-type pp-wave solutions in higher dimensions
\cite{ELEPP,WAVE}, which were found for a class of electrically
charged black holes which preserve $1\over 2$ of supersymmetries
\cite{WAVE,BEHRNDT2}.  It is of interest to find out how the
dyonic BPS saturated solutions, which preserve $1\over 4$ of $N=4$
supersymmetry, are related to the higher-dimensional p-brane
solutions and/or pp-wave solutions.

\acknowledgments
The work is supported in part by U.S. Department of Energy Grant No.
DOE-EY-76-02-3071, the National Science Foundation Grant No. PHY94-07194,
the NATO collaborative research grant CGR 940870 and the National Science
Foundation Career Advancement Award PHY95-12732.

\end{document}